\documentstyle[aps,prl,preprint]{revtex}
\begin{document}
\draft
\title{Effect of gravitational radiation reaction on \\
	circular orbits around a spinning black hole }
\author{Fintan D. Ryan}
\address{Theoretical Astrophysics, California Institute of Technology,
	Pasadena, California 91125}
\date{\today}
\maketitle
\begin{abstract}

The  effect of  gravitational radiation  reaction  on circular  orbits
around  a spinning (Kerr) black hole  is computed to  leading order in
$S$ (the magnitude  of the spin angular momentum  of the hole)  and in
the strength of  gravity  $M/r$ (where $M$   is the mass of  the black
hole, $r$ is the orbital radius, and $G=c=1$).  The radiation reaction
makes the orbit shrink but leaves  it circular, and drives the orbital
plane very  slowly  toward antialignment with  the  spin of  the hole:
$\tan (\iota /2) = \tan  (\iota_0 /2) [1+(61/72)(S/M^2) (M/r)^{3/2}]$,
where $\iota$ is the angle between the normal to the orbital plane and
the  spin direction, and $\iota_0$   is the initial  value of $\iota$,
when $r$ is very large.

\end{abstract}

\narrowtext

The earth-based  LIGO/VIRGO network  of gravitational   wave detectors
(which is now under construction) will be used to search for and study
the gravitational waves from ``particles'', such  as neutron stars and
small black holes, spiraling into massive black  holes (mass $M$ up to
$\sim  300 M_\odot$); and   ESA's planned space-based LISA~\cite{lisa}
interferometer will  do the same for  inspiral into supermassive black
holes ($M$ up  to $\sim  10^7 M_\odot$).  To  search  for the inspiral
waves  and extract the information they   carry will require templates
based  on theoretical  calculations of the   emitted waveforms; and to
compute the waveforms    requires a  detailed understanding of     how
radiation reaction influences the orbital evolution.

For several  years a stumbling  block has impeded  computations of the
evolution, when the orbital  plane of the  particle is inclined to the
equatorial plane  of  a spinning hole:  No practical  method has  been
developed to deduce how radiation reaction influences the evolution of
the orbit's ``Carter  constant''~\cite{carter,mtw}, which governs  the
orbital shape and inclination angle.  This  {\it Letter} describes the
first progress  on  this problem: a  ``post-Newtonian''  gravitational
radiation reaction force is used to compute the full orbital evolution
to first order in $S$, the  magnitude of the  spin angular momentum of
the black hole, and leading order in the  strength of gravity $M/r$ at
the orbital radius $r$.  (Here and  throughout, units with $G=c=1$ are
used.)    The analysis is   restricted to  orbits  that initially  are
``circular''  (more  precisely,  orbits  which have    constant radius
$r$---these orbits  are circular in   the ``orbital plane''  discussed
below, but this plane precesses).  However, the  method can readily be
extended to noncircular orbits~\cite{ecc}  and (with considerably more
difficulty) should  be extendible to the  fully relativistic regime $r
\sim M$.

The  computation of the evolution  presented here proceeds as follows:
First,  in the absence of radiation  reaction, the  orbital motion and
the associated constants  of motion are  reviewed.  Then,  the leading
order  radiation reaction  accelerations   that  act on  the  orbiting
particle  and  on  the hole  are  derived  and   used to  compute  the
radiation-reaction-induced   evolution   of the  constants  of motion.
Finally,  the evolution   of  the  orbit---its shape  and  inclination
angle---is obtained.

The  leading  order effect  of  the spin on  the (otherwise Newtonian)
orbit was  deduced long ago by  Lense and Thirring \cite{lt} (reviewed
by Landau and  Lifshitz~\cite{ll})---though, of  course, they regarded
the  central body as  a star rather  than a black  hole.  In fact, our
analysis does not require the body to be a  black hole, but since this
is the primary case of physical interest, the discussion is phrased in
terms of a black hole.

Let spherical polar coordinates, $r$,  $\theta$, and $\phi$,  centered
on the black  hole, be used  to describe the  location of the particle
(these  coordinates describe  the    relative separation of  the   two
bodies),  with     the  hole's  spin along   the     polar axis.   The
Lagrangian~\cite{kww} for the motion of the  particle (which, for now,
does  not have to be  circular) is given, to   linear order in $S$ but
otherwise in solely Newtonian theory, by
\begin{equation}
\label{eq:lagr}
{\cal L}  =
 \frac{\mu}{2} \left(\dot  r^2 + r^2 \dot  \theta^2 + r^2 \sin^2
	(\theta) \dot \phi^2\right)
 + \frac{\mu M}{r} - \frac  {2 \mu S \sin^2 \theta}{r} \dot \phi,
\end{equation}
where   $\mu$ is the  mass  of the  particle.   In general, an overdot
represents $d/dt$.  The entire analysis is to  leading order in $\mu$.
To leading  order in $S$ and in  $M/r$, the motion resulting from this
Lagrangian  is the  same as  in the Kerr  metric,  which describes the
gravitational   field  of a  spinning hole.   The   use of  flat space
coordinates, which ignores  $M/r$ corrections, is  adequate to leading
order.

Following standard procedure \cite{mech}, the Hamilton-Jacobi equation
associated with the Lagrangian~(\ref{eq:lagr}) can  be shown to have a
separation-of-variables $(t, \phi, r, \theta)$ solution, which reveals
three constants of motion:
\begin{mathletters}
\label{eq:const}
\begin{eqnarray}
E=&&\frac{\mu}{2}\left(\dot r^2 +r^2 \dot \theta^2 + r^2 \sin^2 (\theta) \dot
  \phi^2 \right) - \frac{\mu M}{r},\\
L_z =&& \mu r^2 \sin^2(\theta)\dot \phi -\frac{2 \mu S \sin^2 \theta}{r},\\
Q+L_z^2=&&\mu^2r^4\left(\dot \theta^2 + \sin^2 (\theta) \dot \phi^2\right)
  - 4 \mu^2 S r \sin^2 (\theta) \dot \phi.
\end{eqnarray}
Comparing these expressions  to Eqs.~(33.31) of Ref.~\cite{mtw}, it is
clear that $E$, $L_z$, and $Q$ correspond, to leading order in $S$ and
in $M/r$, to the constants of  motion for a test  particle in the Kerr
metric:  the  energy minus the  test  particle mass, the $z$-component
(component along the  spin axis of the  hole) of angular momentum, and
the Carter constant, respectively.
\end{mathletters}

The     analysis  of radiation   reaction,    below,  is in  Cartesian
coordinates, $x_1= r\sin \theta \cos \phi$, $x_2= r\sin \theta \sin
\phi$, and $x_3 = r \cos \theta$.  In these coordinates, the constants
of motion (\ref{eq:const}) become (repeated indices are summed)

\begin{mathletters}
\label{eq:constincart}
\begin{eqnarray}
E=&&\frac{\mu}{2}\dot x_j \dot x_j-\frac{\mu M}{\left(x_j x_j\right)^{1/2}},\\
L_z=&&\mu \epsilon_{3jk} x_j \dot x_k
	- \frac{2 \mu S \left(x_1^2+x_2^2\right)}
	{\left(x_jx_j\right)^{3/2}},\\
Q+L_z^2 = &&\mu^2 \epsilon_{ijk}x_j \dot x_k\epsilon_{ilm}x_l \dot x_m
  - \frac{4\mu^2 S \epsilon_{3jk}x_j \dot x_k}{\left(x_m x_m\right)^{1/2}}.
\end{eqnarray}
\end{mathletters}

Let the orbital plane have inclination angle $\iota$ (restricted to $0
\leq \iota \leq
\pi$, where  $\iota  >   \pi/2$ corresponds to    an  orbit counter-rotating
relative to the spin), defined as
\begin{equation}
\label{eq:thetainc}
\cos \iota \equiv \frac{L_z}{\left(Q+L_z^2\right)^{1/2}}.
\end{equation}

The constants of motion admit  orbits of constant  radius, just as for
the   Kerr metric  in    Boyer-Lindquist   coordinates.  One  of   the
Euler-Lagrange equations  implies   that, for  $r$  to  be   constant,
$\partial {\cal L} / \partial r = 0$, which leads to
\begin{equation}
\label{eq:radius}
r=M v^{-2} \left[1 + 6 (S/M^2) v^3 \cos \iota \right],
\end{equation}
\begin{equation}
\label{eq:defofv}
v \equiv \left(\frac{Q+L_z^2}{\mu^2 M^2} \right)^{-1/2}.
\end{equation}
The positive root in any square-root is always chosen.

The   following relationship between  the   constants  of  motion  for
circular orbits is easily derived:
\begin{equation}
\label{eq:circtocirc}
E=- \case 1/2 \mu v^2 \left[1-4(S/M^2) v^3 \cos \iota \right].
\end{equation}

The other two   Euler-Lagrange equations predict that  $\theta(t)$ and
$\phi(t)$ are the same as for a circular orbit in the case when $S=0$,
except that $\dot \phi$ is altered to $\dot \phi = \dot \phi |_{S=0} +
2   S /r^3$.   By   transforming these  angular  motions to  Cartesian
coordinates, the following equations for the orbit are obtained:
\begin{mathletters}
\label{eq:x}
\begin{eqnarray}
x_1 = &&r\biggr[
\cos \left(\Omega_{\theta} t\right) \cos \left(\frac{2St}{r^3}\right)
  \cos \iota
  -\sin \left(\Omega_{\theta} t\right)
  \sin \left(\frac{2St}{r^3}\right)\biggr],\\
x_2 = &&r\biggr[
\sin \left(\Omega_{\theta} t\right)\cos \left(\frac{2St}{r^3}\right)
  +\cos \left(\Omega_{\theta} t\right)
  \sin \left(\frac{2St}{r^3}\right)\cos \iota \biggr],\\
x_3 = &&r \cos \left(\Omega_{\theta}t\right) \sin \iota,
\end{eqnarray}
where the angular velocity in the $x_3$ direction is
\end{mathletters}
\begin{equation}
\Omega_{\theta} \equiv M^{-1} v^3 \left[1 - 12 (S/M^2) v^3 \cos \iota \right].
\end{equation}
These  equations  describe a  circular  orbit on a  plane (the orbital
plane, which  is  inclined at angle  $\iota$ to  the equatorial plane)
which precesses  around the spin  axis of the  black hole with angular
velocity $2S/r^3$  (the    Lense-Thirring precession).   Because   the
particle's motion is the the sum of the  circular motion and the plane
precession, the particle itself does not travel  on a fixed plane, nor
does it travel in a circle, nor does  it cross the equatorial plane at
angle $\iota$, but rather at an angle $\iota'$~\cite{inc}.

Turn,  now, to   the  gravitational   radiation   reaction force   (or
acceleration)   that slowly modifies   the above  orbital motion.  The
emitted waves and their associated radiation reaction can be expressed
in terms of  the multipole moments  of  the ``system'' (particle  plus
hole) \cite{th80,bad}.  In  the absence of  spin, $S=0$, the reaction,
at  leading  order in  $M/r$, is due   to  the mass  quadrupole moment
$I_{ij}$  of the system; when $S  \neq 0$, the leading order influence
of $S$  on the reaction is due  in part to  the mass quadrupole moment
$I_{ij}$ and  in part to the  current quadrupole moment $J_{ij}$.  The
moments  $I_{ij}$ and $J_{ij}$ are  usually  written as integrals over
the  mass and momentum densities  of the source.   For the black hole,
however, this cannot be done, and even for a neutron star the standard
integrals are   invalid because they  ignore   the star's relativistic
gravity.   There is an  alternative approach,  however, that does work
for this  black-hole-plus-particle system: the  moments are defined in
terms of the weak,  asymptotic gravitational fields  far from the hole
and particle.   When this is  done, all the  standard formulas  of the
multipolar gravitational wave formalism remain valid \cite{th80}.

The standard theory  of   the radiation reaction  force  (for example,
Ref.~\cite{bad})   is  generally  formulated   in   the center-of-mass
Cartesian  coordinate  system    $\{x_j'\}$, which  differs   from the
black-hole-centered  coordinates   $\{x_j\}$  used   above.    In  the
asymptotic, center-of-mass    coordinates, the hole    (or rather  its
asymptotically spherical gravitational field \cite{th80,thornehartle})
moves along the  path $x_k'(t) =  -(\mu/M)x_k(t)$ with its  spin still
pointing in the $x_3$ direction, and the particle moves along the path
$x_k'(t) = [1-(\mu/M)]x_k(t)$.   Correspondingly, to  leading order in
$\mu/M$, the mass and current quadrupole moments are
\begin{mathletters}
\label{eq:ij}
\begin{eqnarray}
\label{eq:iij}
I_{ij} =&&\left[\mu x_{i} x_{j}\right]^{STF},\\
\label{eq:jij}
J_{ij} =&&\left[\mu x_{i}  \epsilon_{jkm}x_{k} \dot x_{m} -  \case 3/2 (\mu/M)
  x_{i} S \delta_{j3} \right]^{STF}.
\end{eqnarray}
Here, STF means ``symmetrize and remove the trace'', and $x_i = x_i(t)$
is the trajectory of the particle in the hole-centered coordinates, as
given by Eqs.\ (\ref{eq:x}).
\end{mathletters}

Eqs.~(\ref{eq:ij}) agree with  the $\mu  \ll M$  limit of  the moments
given in  Kidder,  Will, and Wiseman  (Ref.~\cite{kww}, Eqs.~(14a,c)).
The first  term of $J_{ij}$  is the standard  contribution  due to the
motion of the particle; the second term  arises from the motion of the
spin of the hole relative  to the asymptotic, center-of-mass  inertial
frame:  when the current
dipole  moment,  due to the   hole's spin  angular   momentum $J_i = S
\delta_{i3}$, is displaced from the system's center of mass by $\delta
x_i' =  -(\mu/M) x_i$ due  to  the orbital  motion, that  displacement
produces a current quadrupole moment  $J_{ij} = [\case 3/2 \delta x_i'
J_j]^{STF} = [$second term  in expression (\ref{eq:jij})], as one  can
deduce from the asymptotic metric  components by which the moments are
defined: Eq.~(11.1b) of Ref.~\cite{th80}.

Consider, for  the   moment,  the  radiation  reaction force    acting
individually on  the  hole or  on   the particle.  Let the   object of
interest (hole or particle) have  a location $x_j'$ and velocity $\dot
x_j'$ in the asymptotic,   center-of-mass coordinates.  Then, for  the
moment ignoring the object's spin, its radiation reaction acceleration
(force divided by mass) is given by
\begin{eqnarray}
\label{eq:geo}
a^{(react)}_j =
&&- \case 2/5 I^{(5)}_{jk}x_k'+\case{16}/{45}\epsilon_{jpq}
   J^{(6)}_{pk} x_q' x_k'
  +\case{32}/{45} \epsilon_{jpq} J^{(5)}_{pk} x_k' \dot x_q'
  +\case{32}/{45} \epsilon_{pq[j}J^{(5)}_{k]p}x_q' \dot x_k'.
\end{eqnarray}
This   can  be  derived  from  Eqs.~(11)   and   (12) of Blanchet  and
Damour~\cite{bad},  keeping   only  the mass  quadrupole   and current
quadrupole     moment     terms.     Here,     the  brackets    denote
antisymmetrization: $C_{[jk]} \equiv (C_{jk} - C_{kj})/2$.  The number
in  parentheses to the upper   right of a  multipole moment  indicates
taking that  number of  time  derivatives.

The   coupling  of the  object's spin   angular  momentum $J_i$ to the
radiation reaction  field  due   to $J_{ij}$ produces    an additional
radiation reaction acceleration,
\begin{equation}
\label{eq:bforce}
a_j^{(react)}|_{spin} = - \case 8/{15} J_{ij}^{(5)} J_i/m,
\end{equation}
where $m$  is the  mass of the  object.   If the spinning  object were
nearly Newtonian, Eq.~(\ref{eq:bforce}) could be  derived by adding up
the  velocity  dependent radiation   reaction  force [mass$\times$Eq.\
(\ref{eq:geo})] on each    bit of mass   inside the   source  and then
dividing by the total  mass $m$ of  the object.   For the  black hole,
such a procedure   is invalid; however, the  result  (\ref{eq:bforce})
must  still   be   true:    The   analysis of    Thorne    and  Hartle
\cite{thornehartle}, specifically their   Eq.~(1.9b), shows  that  the
force on any isolated  spinning object in a ``gravitomagnetic  field''
(the  type of radiation  reaction field  that  is responsible for this
force \cite{bad}) is  the same as for  a spinning mass with only  weak
self-gravity.

Since the constants of  motion, $E$, $L_z$, and  $Q$, are  defined and
expressed in  terms of   black-hole-centered coordinates   rather than
center-of-mass coordinates, their evolution must be computed using the
radiation reaction acceleration of the particle  relative to the black
hole,  i.e.\ the difference  of the accelerations  of the particle and
the hole relative to the center-of-mass:
\begin{mathletters}
\label{eq:atot}
\begin{eqnarray}
a_j &&\equiv a^{(react)}_j|_{particle} - a^{(react)}_j|_{black~hole}\\
\label{eq:atotal}
&&= - \case {2}/{5}I^{(5)}_{jk}x_k+\case{16}/{45}\epsilon_{jpq}
   J^{(6)}_{pk} x_q x_k +\case{32}/{45} \epsilon_{jpq}J^{(5)}_{pk}x_k \dot x_q
   +\case{32}/{45} \epsilon_{pq[j}J^{(5)}_{k]p}x_q \dot x_k+
   \case{8}/{15} (S/M) J_{3j}^{(5)}.
\end{eqnarray}
Here,   the    moments  $I_{ij}$     and    $J_{ij}$  are   given   by
Eqs.~(\ref{eq:ij}).  There is no contribution to (\ref{eq:atotal}) (up
to  leading order in  $\mu$) from the  acceleration on the hole except
through the spin interaction (\ref{eq:bforce}),  which gives the fifth
term in Eq.~(\ref{eq:atotal}).  For the  $J_{ij}$ in the  contribution
to the  acceleration on the  particle  [the second,  third, and fourth
terms  of  Eq.~(\ref{eq:atotal})],     only   the  second    term   of
Eq.~(\ref{eq:jij}) needs to be  kept: the contribution from  the first
term is not at leading  order in $M/r$ for  either terms not involving
$S$ or terms involving $S$.  Similarly, the $J_{3j}$ in the fifth term
of Eq.~(\ref{eq:atotal}) requires only the contribution from the first
term of Eq.~(\ref{eq:jij}).
\end{mathletters}

By   differentiating Eqs.~(\ref{eq:constincart}) with  respect to time
and using the $a_j$  of Eqs.~(\ref{eq:atot}) as the radiation reaction
contribution to $\ddot x_j$, we obtain the  following evolution of the
``constants'' of motion:
\begin{mathletters}
\label{eq:elqdot}
\begin{eqnarray}
\dot E = &&\mu \dot x_ja_j,\\
\dot L_z = &&\mu \epsilon_{3jk}x_j a_k,\\
\dot {Q+L_z^2} = && 2 \mu^2 \epsilon_{ijk}x_j
   \dot x_k\epsilon_{ilm}x_la_m
  - \frac{4 \mu^2 S \epsilon_{3jk}x_j a_k}{\left(x_m x_m\right)^{1/2}}.
\end{eqnarray}
\end{mathletters}

While Eqs.~(\ref{eq:ij})--(\ref{eq:elqdot}) are valid for  noncircular
orbits as well as circular orbits, we will  treat only circular orbits
here.  Suppose that the orbit initially is circular, so the trajectory
of the particle $x_k(t)$ in the  hole-centered coordinates is given by
Eqs.~(\ref{eq:x}).   Then, the evolution  of  its constants of motion,
$E$, $L_z$,  and $Q+L_z^2$,  can be  computed  as follows.  Insert the
trajectory of the particle (\ref{eq:x}) into Eqs.~(\ref{eq:ij}) to get
the $I_{ij}$ and $J_{ij}$ moments.  Then, insert these moments and the
trajectory of  the   particle into  Eq.~(\ref{eq:atotal}) to  get  the
relative radiation reaction   acceleration.  Insert this  acceleration
into Eqs.~(\ref{eq:elqdot}) for  the  time derivatives of $E$,  $L_z$,
and $Q+L_z^2$,  and average the resulting   expressions over an orbit.
[The averaging, denoted by $\langle~\rangle$, can  be taken to involve
times  $-\pi/\Omega_\theta < t <   +\pi/\Omega_\theta$, and because of
this   restriction     to   small      $t$,  for    terms      in  the
trajectory~(\ref{eq:x})  that have argument  $2St/r^3$, the cosine can
be  replaced by  $1$ and the  sine  can  be  replace by its  argument,
thereby    simplifying      the   calculation.]         The     result
is~\cite{mathematica}
\begin{mathletters}
\label{eq:adot}
\begin{eqnarray}
\label{eq:aedot}
\Bigl\langle \dot{E}\Bigr\rangle = -&&\frac{32}{5} \frac{\mu^2}{M^2} v^{10}
\left(1-\frac{433}{12} \frac{S v^3} {M^2} \cos \iota \right),\\
\label{eq:aldot}
\Bigl\langle \dot{L_z}\Bigr\rangle= -&&\frac{32}{5} \frac{\mu^2}{M} v^7
\left(\cos \iota +\frac{61-687 \cos^2 \iota}{24} \frac{S v^3} {M^2}\right),\\
\label{eq:aqldot}
\Bigl\langle \dot{Q+L_z^2}\Bigr\rangle = -&&\frac{64}{5} \mu^3 v^6
\left(1-\frac{313}{12} \frac{Sv^3} {M^2} \cos \iota \right).
\end{eqnarray}
Eqs.~(\ref{eq:aedot})   and   (\ref{eq:aldot})  agree  (after  trivial
conversions  of notation) with  previous results (Ref.~\cite{shibata},
Eqs.~(3.13),  (3.18)  and    Ref.~\cite{kidder}, Eqs.~(4.10), (4.11)),
computed   by    alternative  methods  that cannot    compute $\langle
\dot{Q+L_z^2} \rangle$.
\end{mathletters}

These time  derivatives of  the constants  of  motion have  two  major
implications for the orbital evolution:   First, they imply that   the
orbit, which was   assumed initially circular, remains circular;  this
can be seen   from  the fact  that  they preserve   the circular-orbit
relationship~(\ref{eq:circtocirc}).       Second,  when combined  with
Eqs.~(\ref{eq:thetainc}), they  imply  the following evolution  of the
orbital plane's inclination angle:
\begin{equation}
\label{eq:iotadot}
\biggr\langle \frac{d \iota}{dt} \biggr\rangle
= \frac{244}{15} \frac{\mu S}{M^4} v^{11} \sin \iota.
\end{equation}
This equation implies that radiation reaction drives the orbital plane
toward  antialignment  with the spin,   as {\it might}  be intuitively
expected  since that orientation  minimizes  the energy of  spin-orbit
coupling~\cite{intu}.

By combining Eq.~(\ref{eq:iotadot})  with the leading  order change in
$r$, $\dot  r =  -{64\over5}(\mu/M)  v^6$, the  following differential
equation relating $r$ and $\iota$ is obtained:
\begin{equation}
\label{eq:diffeq}
\frac{d \iota}{- d \ln r}
= \frac{61}{48} \frac{S}{M^2} \left(\frac{M}{r} \right)^{3/2} \sin \iota.
\end{equation}
The time-averaging of  the left hand  side of Eq.~(\ref{eq:diffeq}) is
not explicitly written down.  Integrating Eq.~(\ref{eq:diffeq}) yields
\begin{equation}
\label{eq:evol}
\tan (\iota /2)
= \tan  (\iota_0  /2) [1+\case{61}/{72}  (S/ M^2) (M/r)^{3/2}],
\end{equation}
where  $\iota_0$ is the value of  the inclination angle when the orbit
has     large radius.

Eq.~(\ref{eq:evol}) shows, for  example, that for  an $S \lesssim M^2$
black hole, if the particle orbits with  a small inclination angle $(0
\lesssim \iota \ll \pi)$, then  $\iota$ fractionally changes by $0.06$
from its initial value at large radius to its value at $r \approx 6M$.
The regime $r  \lesssim 6M$ is of special  interest; it is  there that
waves from the final stages of inspiral can give high-accuracy maps of
the hole's spacetime  geometry~\cite{rft}.  However, when $r \approx 6
M$, our leading order  analysis breaks down,  so that to be able
to map the hole's geometry, the analysis must be carried out
to higher order in $M/r$ and $S$.

The  above analysis  illustrates the  power  of the radiation reaction
force method  to  reveal the   detailed  evolution of  a system  under
radiation  reaction.  The case solved  was sufficiently simple to give
an   easily presented solution.   A  future paper~\cite{ecc} will give
more details of the above  calculations, along with the generalization
to  eccentric orbits.  Hopefully, future  work with radiation reaction
forces  will: (1)~Generalize the  analysis to an  arbitrary mass ratio
$\mu/M$ and  the case of  both masses   having spin.   (2)~Extend  the
analysis to  higher order in $M/r$ and  in $S$.  (3)~Achieve a similar
calculation of  the orbital evolution  in the  fully relativistic Kerr
metric.

The author is grateful to Alan Wiseman and Kip  Thorne for their advice.
This work was supported by NSF grants AST-9114925 and AST-9417371, and
by NASA grants NAGW-2897 and NAGW-4268.


\begin{references}
\bibitem{inc} The angle $\iota'$ is defined as  the angle between the
Newtonian orbital angular momentum  $L_{N_i} \equiv \mu \epsilon_{ijk}
x_j \dot x_k$ and the spin axis, so that when the particle crosses the
equator, it does so at angle $\iota'$.  However, neither $L_{N_i}$ nor
$\iota'$  is a constant of the  motion.  For circular orbits, $\iota'$
is given by $\cos \iota' = \cos  \iota + 2  (S/M^2) v^3 (\sin^2 \theta
-\cos^2 \iota)$.
\bibitem{intu}  Other definitions  of ``inclination  angle'', such as
$\iota'$ at $\theta = \pi/2$ (see~\cite{inc}), can  lead to such angle
decreasing with time.
\end{references}
\end{document}